\documentclass[a4paper]{article}
\usepackage{amsmath}
\usepackage[english]{babel}
\usepackage{latexsym}
\usepackage{amssymb}
\usepackage{amscd}
\usepackage{amsgen,amstext,amsbsy,amsopn}
\usepackage{math rsfs}

\numberwithin{equation}{section}
\newcommand{\bdm}{\begin{displaymath}}
\newcommand{\edm}{\end{displaymath}}
\newcommand{\bdn}{\begin{eqnarray}}
\newcommand{\edn}{\end{eqnarray}}
\newcommand{\bay}{\begin{array}{c}}
\newcommand{\eay}{\end{array}}
\newcommand{\ben}{\begin{enumerate}}
\newcommand{\een}{\end{enumerate}}
\newcommand{\beq}{\begin{equation}}
\newcommand{\eeq}{\end{equation}}

\newcommand{\R}{\mathbb{R}}

\newcommand{\RT}{\mathbb{R}^3}

\newcommand{\xvp}{\xv^{\prime}}
\newcommand{\rv}{\mathbf r}

\newcommand{\diff}{\mathrm{d}}
\newcommand{\eps}{\varepsilon}

\newcommand{\gpf}{\mathcal{E}^{\mathrm{GP}}}

\newcommand{\gpe}{E^{\mathrm{GP}}}

\newcommand{\tff}{\mathcal{E}^{\mathrm{TF}}}

\newcommand{\ttfe}{E^{\mathrm{TF}}_{1,\omega}}

\newcommand{\tfeinf}{E^{\mathrm{TF}}_{0,1}}

\newcommand{\half}{\hbox{$\frac12$}}

\newcommand{\E}{\mathcal{E}}
\newcommand{\xv}{\mathbf {x}}
\def\dint{\mathop{\displaystyle \int}}%

\newtheorem{theorem}{Theorem}[section]

\begin{document}

\markboth{Jakob Yngvason}
{Bosons in Rapid Rotation}

%
%

\title{Bosons in Rapid Rotation\footnote{Lecture given at a symposium in honor of Michel Dubois-Violette, Orsay,  April 2008}}

\author{Jakob Yngvason	\\ \normalsize\it Fakult\"at f\"ur Physik, Universit{\"a}t Wien,	\\ \normalsize\it Boltzmanngasse 5, 1090 Vienna, Austria, \\ \normalsize\it Erwin Schr{\"o}dinger Institute for Mathematical Physics,	\\ \normalsize\it Boltzmanngasse 9, 1090 Vienna, Austria.}


\maketitle


\begin{abstract}
Some recent progress in the mathematical  physics of rapidly rotating,  dilute Bose gases in anharmonic traps is reviewed.
\end{abstract}


\section{Introduction}	

Since the first experimental realization of Bose-Einstein condensation in trapped, dilute alkali gases in 1995  the strange quantum properties exhibited
by ultra\-cold quantum gases have become a large research field worldwide, both experimentally and theoretically. Under  normal conditions the atoms of a dilute gas are distributed on very many one-particle quantum states so that every single state is only occupied by relatively few atoms. By cooling a gas of alkali atoms to extremely low temperatures (of the order $10^{-8}$ K) it is, however,  possible to achieve Bose-Einstein condensation, where a \emph {single} quantum state is, on the average, occupied by a \emph{macroscopic} number of particles. This phenomenon was predicted by Albert Einstein in 1924  based on a model of an \emph {ideal} quantum gas, where the unavoidable interactions between the atoms are ignored. The interactions, however, are important and modify the phenomenon in several ways. Despite much progress in the theoretical understanding (see, e.g.,  the monographs \cite{PiSt2003} and \cite{PS}) it is still a real challenge for mathematical physics to derive the properties of the low energy states of interacting Bose gases from the many-body Hamiltonian by rigorous mathematical analysis. I refer to  \cite{LSSY} for an introduction to these problems.

In the present contribution the focus will be the effects of \emph {rotations} on the ground state properties of dilute Bose gases. In particular, I shall review some recent results obtained in collaborations with \emph{Jean-Bernard Bru, Michele Correggi, Tanja Rindler-Daller and Peter Pickl} \cite{CDY1, CDY2, BCPY, CY}. These results concern mainly the ground state energy of a gas in a rapidly rotating container as well as the vortices that are created by the rotation  and arrange themselves in a geometric pattern to minimize the energy. The external potential modelling the container is here assumed to increase faster than quadratically with the distance from the rotational axis (\lq anharmonic traps'), allowing arbitrarily high rotational velocities.  By contrast, in a quadratic confining potential (\lq harmonic trap') the centrifugal forces will drive the gas out of the container when the rotational velocity exceeds a finite value. When this limiting velocity is approached from below interesting new effects are expected to occur but they are not subject of the present review except for some brief comments in Section \ref{harmonic}. 

\section{The General Setting}

We consider  the quantum mechanical Hamiltonian for $N$ 
bosons with a pair interaction potential, $v$, and  external potential,
{$V$}, in a rotating frame with angular velocity ${\mathbf \Omega}$:
\begin{equation}
H = 
\sum_{j=1}^{N} \left(- \nabla^2_j +V({\mathbf x}_{j})-{\mathbf L}_j\cdot {\mathbf \Omega}\right)+
\sum_{1 \leq i < j \leq N} v(|{\mathbf x}_i - {\mathbf x}_j|).
\end{equation}
Here  ${\mathbf x}_j\in\R^3$, $j=1,\dots,N$ are the positions and 
${\mathbf L}_j=-\hbox{i}\,{\mathbf x}_j\wedge{\mathbf \nabla}_j$ the angular momentum operators of the
particles. Units have been chosen so that $\hbar=2m=1$. 

The pair interaction potential $v$ is assumed to be radially symmetric and of finite range. For the mathematical results in the sequel we also assume that $v$ is nonnegative.  The Hamiltonian  operates on {\it symmetric} (bosonic) wave functions
in $L^2(\mathbb R^{3N}, \diff^3{\mathbf x}_1\cdots \diff^3{\mathbf x}_N)$.

The Hamiltonian can alternatively  be written in the form 
\begin{equation}\label{ham}
H= 
\sum_{j=1}^{N} \left((-{\rm i}\nabla_j-{\mathbf A}({{\mathbf x}}_j))^2+V({\mathbf x}_{j})-\hbox{$\frac 14$}\Omega^2r_j^2\right)+
\sum_{1 \leq i < j \leq N} v(|{\mathbf x}_i - {\mathbf x}_j|)
\end{equation}
\noindent with ${\mathbf A}({\mathbf x})=\hbox{$\frac12$}{\mathbf \Omega}\wedge {\mathbf x}$ and $r$ the distance from the rotation axis. 
This corresponds to a splitting of the rotational effects into Coriolis and centrifugal forces.
The Coriolis forces have formally the same effect as a  magnetic field with vector potential ${\mathbf A}({\mathbf x})$.

We are interested in the ground state properties of the Hamiltonian for $N\to \infty$, in particular the effect of 
rotations on  dilute gases. Dilute means that
\beq\label{dilute} a^3\rho\ll 1\eeq   where $a$ is the \emph{scattering length} of $v$ and $ \rho$ an appropriately defined average particle density.
The scattering length is defined by considering the
zero energy scattering equation 
\beq -\nabla^2\psi+\half v\psi=0.\eeq  
For $|{\mathbf x}|$ larger than the range of $v$ the solution is
\beq \psi({\mathbf x})=({\rm const.})\left(1- a{|{\mathbf x}|^{-1}}\right)\eeq
which defines $a$.
 The condition $a^3\rho\ll 1$ can also be written
\beq a\ll \rho^{-1/3}\eeq  i.e., the scattering length is much smaller than the mean particle distance.

A special case of a dilute limit  is the  Gross-Pitaevskii (GP) limit, which means that  the GP coupling parameter
\beq g\equiv4\pi\, aN/L\eeq  
is kept constant as $N\to\infty$. Here  $L$ is the length scale associated with $-\nabla^2+V$, measured, e.g., in terms of the width $\Delta \bf x$ of the ground state of this operator.
Note that $\rho\sim N/L^3$, so  in the GP limit
\beq a^3 \rho\sim {g^3}/{N^2}=O(N^{-2}).\eeq  
 
    To understand the meaning of the GP limit one should recall a basic formula for the ground state energy per particle of a
   dilute  Bose gas in a box:
    \beq \label{gse}E_0/N\approx 4\pi \rho a.\eeq  
 Thus, in the GP limit, the interaction energy per particle,  $\sim Na/L^3$, is of the same order as the lowest excitation energy without interaction,  $\sim 1/L^2$.
   
   The formula (\ref{gse}) has an interesting history, spanning almost 80 years! The technique used for the first complete mathematical proof \cite{LY1998} (see also \cite{LY2d} for the two-dimensional case) has been important for subsequent  rigorous work on dilute Bose gases.

     Associated with the energy $\sim\rho a$ there is a natural length scale, the  \lq healing length'
     \beq \ell_c=(4\pi \rho a)^{-1/2}.\eeq  
Thus  $1/\ell_c^2$ is equal to the interaction energy per particle.
If the density of an inhomogeneous gas is somewhere zero, due to boundary conditions or the formation of vortices, the interaction will tend to restore the full density (\lq heal') on a length scale of order $\ell_c$. Note that $\ell_c$  decreases if the scattering length increases.
For later use it is convenient to define
\beq \varepsilon\equiv \ell_c/L=g^{-1/2}.\eeq  
In rotating gases  $\varepsilon$ is a measure for the size of the vortex cores relative to the size of the trap. 

Besides the energy it is also of interest to consider the
\emph{particle density}  
\beq\rho_{0}({\mathbf x})=N\int_{\R^{3(N-1})}|\Psi_{0}({\mathbf x},{\mathbf x}_{2},\dots,{\mathbf x}_{N})|^2\diff^3{\mathbf x}_2\cdots
\diff^3{\mathbf x}_N\eeq associated with a (ground state) wave function $\Psi_{0}$, and the \emph{one-particle density matrix} 
\begin{equation}
\gamma_0({\mathbf x},{\mathbf x}')= N \int_{\R^{3(N-1})} \Psi_0({\mathbf x},{\mathbf x}_2,\dots,{\mathbf x}_N) 
\Psi_0^{\ast}({\mathbf x}',{\mathbf x}_2,\dots,{\mathbf x}_N)\, \diff^3{\mathbf x}_2\cdots
\diff^3{\mathbf x}_N \ .
\end{equation}
{\it Bose-Einstein condensation} (BEC) in the state $\Psi_0$ means (by definition) 
that the operator defined by the integral kernel 
$\gamma_{0}({\mathbf x},{\mathbf x}')$ has an eigenvalue of the order $N$ for all 
large $N$. This means that   \beq\label{bec} \gamma_{0}({\mathbf x},{\mathbf x}') =\lambda_0\,\psi_0({\mathbf x})\psi_0({\mathbf x}')^*+{\rm Rest}\eeq  
   with $\lambda_0=O(N)$ for large $N$. 
 The ratio $\lambda_0/N$ is called the \emph{condensate fraction}  and $\psi_0$ the \emph{wave function of the condensate} (or sometimes the \lq superfluid order parameter'). The \lq rest' in (\ref{bec}) corresponds to the \emph{depletion of the condensate}, i.e., the contribution of states orthogonal to $\psi_0$.
   
  Using creation and annihilation operators,   we can more generally  write
  \beq \gamma_{0}({\mathbf x},{\mathbf x}')=\langle a^\dagger({\mathbf x})a({\mathbf x}')\rangle\eeq   where $\langle\cdot\rangle$ is the expectation value in the many-body  state that need not be a pure state and could, e.g., be a thermal equilibrium state.   Thus
  \beq \lambda_0=\langle a^\dagger(\psi_0)a(\psi_0)\rangle\eeq  
  is the average particle number in the  one-particle state $\psi_0$
 when the whole system is in the given many-body state.

The following basic results on the GP limit of the ground state of the many-body Hamiltonian for $g=4\pi Na/L$ and $\Omega$ {\it fixed} were established in  \cite{LSY1999, Seir2, Lieb1}:
\begin{itemize}
\item The ground states can be described through minimizers of the \emph{Gross-Pitaevskii energy functional}
\begin{eqnarray}\label{gpfunc}
\E^{\rm
GP}[\psi]&=&\int_{\R^3}\left\{|\nabla\psi|^2+V|\psi|^2-{\bf \Omega}\cdot \psi^*{\bf L}\psi+
g|\psi|^4\right\}\hbox{\rm d}^3{\mathbf x}\nonumber \\
&=&\int_{\R^3}\left\{|({\rm i}\nabla+\,{\mathbf A})\psi|^2+(V-\hbox{$\frac 14$}\Omega^2r^2)|\psi|^2+
g|\psi|^4\right\}\hbox{\rm d}^3{\mathbf x}
\end{eqnarray}
with the normalization condition
{${\int}_{\R^3}|\psi|^2=1$}.
\item For rotating gases the bosonic ground state is in general not the same as the absolute ground state and it need not be unique due to breaking of rotational symmetry.
\end{itemize}

More precisely,  Bose-Einstein condensation takes place in the GP limit and every one particle density matrix, obtained as the limit of  one-particle density matrices, divided by $N$, of ground states of  (\ref{ham}) is a convex combination of projectors onto minimizers of the GP energy functional.

The GP minimizers are solutions of the \emph{GP equation}
\beq \left\{-(\nabla-{\rm i}\,{\mathbf A})^2+(V-\hbox{$\frac 14$}\Omega^2r^2)+2
g|\psi|^2\right\} \psi=\mu \psi.
\eeq  
It has the form of a non-linear Schr\"odinger equation, with a magnetic field in the case of rotation. The main new feature compared to the non-rotating case is the possible occurrence of \emph{vortices}, i.e., singularities of the phase of $\psi$ with integer winding numbers.

The GP equation and its vortex solutions is a subject of its own that can be studied independently of the many-body problem. See  the monograph \cite{Afta} and the review article \cite{FetterRMP} where a large number of references can be found. The most detailed results are for the two-dimensional GP equation, i.e., when $\psi$ depends only on the coordinates in the plane perpendicular to the angular velocity ${\mathbf\Omega}$.
In particular, the  two-dimensional GP equation with a quadratic trap potentials $V(\rv)\sim r^2$ has been studied in \cite{AftaDu, Igna1, Igna2} in the asymptotic parameter regime \beq g\to \infty,\qquad \Omega \sim g^{-1/2}|\log g|.\eeq 
More general, homogeneous trapping potentials are discussed, e.g., in \cite{CDY2, RD}.

\section{Two-Dimensional GP Vortices}

This section  provides some heuristic background for understanding the occurrence of vortices in the case of the two-dimensional GP equation. In order to bring out the salient points as simply as possible we consider a \lq flat'  trap, i.e., the condensate is confined to a circle of radius  $R$ and we impose Neumann conditions at the boundary.

The first thing to note is that for sufficiently  small rotational velocities the condensate stays at \emph{rest in the inertial frame} and thus appears to \emph{rotate opposite to ${\bf \Omega}$ in the rotating frame}. This is a manifestation of superfluidity:  A  normal fluid would pick up the rotational velocity of the container and in equilibrium the fluid would be  at rest in the rotating frame.

In the rotating frame the operator of the  velocity is {$-{\rm i}\nabla-{\bf A}(\bf r)$}. The constant wave function, that minimizes the GP energy functional (with zero energy in excess of the interaction energy) for small $\Omega$, thus has, in the rotating frame,  the velocity
\beq {\mathbf v}(\mathbf r)=-{\bf A({\mathbf r})}=-\half{\mathbf \Omega}\wedge{\mathbf r}=-\half \Omega r \,{ {\mathbf e}_\theta},\eeq  
where ${\mathbf e}_\theta$ denotes the unit vector with respect to the angular variable.
Note that the kinetic energy corresponding to this velocity is exactly compensated by the centrifugal term $-\hbox{$\frac 14$}\Omega^2r^2$ in the GP energy functional (\ref{gpfunc}). 

At higher rotational velocities the condensate responds by creating vortices in order to reduce the velocity.
Consider the case of  {\rm large $g$}, i.e., {{\rm small}  $\varepsilon\sim g^{-1/2}$}. A  \emph{vortex of degree $d$}  located at the origin has the form
{\beq \psi(r,\theta)=f(r)\exp({\rm i}\theta d)\eeq  } 
with 
\beq
        	f(r) \sim
        	\left\{
        	\begin{array}{ll}
            		r^d    &   \mbox{\rm if} \:\:\:\: 0\leq r\lesssim \varepsilon R  \\
            		\mbox{}     &   \mbox{} \\
            		R^{-1}     &   \mbox{\rm if} \:\:\:\: \varepsilon R \lesssim r\leq R       	\end{array}
        	\right.
    	\eeq
	Now the component of the velocity in the direction of ${\mathbf  e}_\theta$ is
	\beq \mathbf v(\mathbf r)_\theta=\left(\frac dr-\half \Omega r\right){\mathbf  e}_\theta\ .\eeq  
	The energy (in excess of the interaction energy) is therefore
\begin{equation}\sim R^{-2}\int_{\varepsilon R}^R[(d/r)^2-d\,\Omega]\,r\,dr+O(1)\\=R^{-2}d^2|\log\varepsilon |-\half d\, \Omega+O(1).\end{equation}
Thus  a vortex of degree $d=1$ can be expected to form when
\beq R^{-2}|\log \varepsilon|-\half \Omega<0\eeq  
i.e., 
\beq \Omega>\Omega_{1}\sim R^{-2}|\log\varepsilon|.\eeq  
We also see that $d$ vortices of degree 1, ignoring their interaction, have energy $\sim d(R^{-2}|\log \varepsilon|- \Omega/2)$ while a vortex of degree $d$ has energy $R^{-2}d^2|\log\varepsilon |- d\,\Omega/2$. Hence it is energetically favorable to \lq split' a $d$-vortex into $d$ pieces of 1-vortices, breaking the rotational symmetry. 
    
             These heuristic considerations are confirmed by a detailed analysis for \lq slowly' rotating gases, i.e., 
          $\Omega=O(|\log \varepsilon|)$ \cite{AftaDu, Igna1, Igna2, RD}:
Vortices start to appear for {$\Omega R^2=\pi |\log \varepsilon |$} and for
	\beq\label{vortexcond} |\log \varepsilon |+(d-1)\log|\log\varepsilon|<\Omega R^2/\pi\leq  |\log \varepsilon |+d\log|\log\varepsilon|\eeq there are exactly $d$ vortices of degree 1. 
If the \lq flat' trap is replaced by  a homogeneous trap, $V(r)\sim r^s$, the effective radius $R$ of the condensate is determined by equating the potential energy and the interaction energy, i.e., 
	$R^s\sim \varepsilon^{-2} R^2(R^{-1})^4$, which leads to 
	 \beq R\sim \varepsilon^{-2/(s+2)}\eeq   and thus (\ref{vortexcond}) is replaced by 
	\beq \Omega_{1}\sim \varepsilon^{4/(s+2)}|\log\varepsilon|.\eeq  
	
	In \cite{Seir1} a number of general results about two-dimensional GP vortices are derived that are not limited to the asymptotic regime $\varepsilon\to 0$. The trap potential is assumed to be rotationally symmetric and polynomially bounded; for $V(r)\sim r^2$ one must in addition assume that $\Omega$ is less than a critical velocity, $\Omega_c$,  so that $V(r)-\Omega^2 r^2/4$ is bounded below. Among the results proved in \cite{Seir1} are
 \begin{itemize}
 \item {Instability for large $d$:} For all $0\leq \Omega<\Omega_c$ there is a $d_\Omega$, independent of the coupling constant $g$, so that all vortices with $d\geq d_\Omega$ are unstable.
 \item {Symmetry breaking:} For all  $0\leq \Omega<\Omega_c$ there is a $g_\Omega$, so that $g\geq
 g_\Omega$ implies that no ground state of the GP energy functional is an eigenfunction of angular momentum. 
\end{itemize}
 
 The limit $g\to\infty$, i.e., $\eps\to 0$,  is usually referred to as a \emph {Thomas-Fermi (TF) limit} because the resulting energy functional, which is just the GP functional without the kinetic energy term,  has a formal similarity with the one studied in TF theory for electrons although the physics is quite different.

\section{Rapid Rotation and Vortex Lattices}

This section contains a r\'esum\'e of \cite{CY}.
The setting is again that of two-dimensional GP theory in a \lq flat', two-dimensional trap trap and by simple scaling we may take $R=1$.  If $\Omega=O(|\log\eps|)$  there is a \emph{finite} number of vortices, even as $\eps\to 0$, as mentioned in the previous section. For larger $\Omega$ the number of vortices is \emph{unbounded} as $\eps\to 0$. Of particular interest is the case $\Omega=O(1/\eps)$. 

For the  flat, 2D trap the GP energy functional is
\beq\label{2dgpfunc} \mathcal E^{\rm GP}[\psi]=\int_{|\rv|\leq 1}\left\{|({\rm i}\nabla+{\mathbf A})\psi|^2-\hbox{$\frac 14$}\Omega^2r^2|\psi|^2+
{\eps^{-2}}|\psi|^4\right\}\hbox{\rm d}^2\rv\eeq  
with ${\mathbf A}(\rv)=\half \Omega\,r\,{\mathbf e}_\theta$.
The special significance of the case $\Omega=O(1/\eps)$ can be seen from the GP energy functional:
 The centrifugal term {$-(\Omega^2/4) r^2|\psi(\rv)|^2$} and the interaction term {$(1/\eps^2)|\psi(\rv)|^4$} are then comparable in size.

  The \lq kinetic energy' term {$|({\rm i}\nabla +{\mathbf A}(\rv))\psi(\rv)|^2$} is formally also of order $1/\eps^2$ if 
  $\Omega\sim 1/\eps$. However, it turns out that its contribution to the energy is, in fact, of lower order because  a lattice of vortices  emerges as $\eps\to 0$. The velocity field generated by the vortices compensates partly that generated by ${\mathbf A}(\rv)=\half {\bf \Omega}\wedge \rv$. The precise statement, proved in \cite{CY}, that puts the analysis of \cite{FiB} on a rigorous basis and generalizes previous results of \cite{CDY1}, is as follows:\\
  
\begin{theorem} [Energy asymptotics, 2D GP theory]\label{thm1} Let $E^{\rm GP}$ denote the GP energy, i.e., the minimum of the GP energy functional (\ref{2dgpfunc}). Let $E^{\rm TF}$ denote the minimal energy of the GP functional {\it  without} the kinetic term.  

If $1/\eps\lesssim  \Omega\ll 1/({\eps^2}|\log\eps|)$  then
  \beq\label{bound1} E^{\rm GP}=E^{\rm TF}+\half\Omega|\log\eps|(1+o(1)).\eeq
   
If $|\log\eps|\ll \Omega\ll 1/\eps$, then
 \beq\label{bound2} E^{\rm GP}=E^{\rm TF}+\half \Omega|\log(\eps^2\Omega)|(1+o(1)).\eeq  
\end{theorem}
  \smallskip
  In particular the contribution of the kinetic term is only $O(|\log\eps|/\eps)$ instead of $O(1/\eps^2)$ if $\Omega\sim 1/ \eps$.
  
  The energy $E^{\rm TF}$ is the infimum of the TF functional
  \beq 
  \mathcal E^{\rm TF}[\rho]=\int_{|\rv|\leq 1}\left\{-\hbox{$\frac 14$}\Omega^2r^2\rho+
{\eps^{-2}}\rho^2\right\}\hbox{\rm d}^2\rv\eeq
where the infimum is over all nonnegative $\rho(\rv)$ with $\int\rho=1$. The energy $E^{\rm TF}$ and the (unique) minimizer, $\rho^{\rm TF}$, are easily computed explicitly. Due to the centrifugal forces the density has a parabolic shape for $\Omega\leq \omega_{\rm hole}/\eps$ with $\omega_{\rm hole}=4/\sqrt{\pi}$, but develops a \lq hole' if  $\Omega> \omega_{\rm hole}/\eps$. For $\Omega\gg \omega_{\rm hole}/\eps$ the density is concentrated in a thin layer of thickness $\sim \eps\Omega^{-1}$ at the boundary of the trap.
  
  In the proof of Theorem \ref{thm1} one uses, for the upper bound,  a trial function corresponding to a \emph{lattice of vortices}. Writing points $\rv=(x,y)\in \mathbb R^2$ as complex numbers, $\zeta=x+iy$, the trial function has the form
  \beq\label{ansatz1} \psi(\rv)=f(\rv)\sqrt{\rho(r)}\exp\{{\rm i}\varphi(\rv)\}\eeq  
  where $\rho(r)$ is  a suitably regularized version of the TF density $\rho^{\rm TF}(r)$ and
  \beq \exp\{{\rm i}\varphi(\rv)\}=\prod_i\frac{\zeta-\zeta_i}{|\zeta-\zeta_i|}\eeq  
  where the points $\zeta_i$ form a regular lattice within the trap. The function $f(\rv)$ 
  has a zero of order 1 at each of the points $\zeta_i$ and is constant outside a disc (\lq vortex core') of radius $t\ll\Omega^{-1/2}$ around each lattice point.
  
  Note that \beq \varphi(\rv)=\sum_i\arg(\zeta-\zeta_i)\eeq  
  is a harmonic function. Its conjugate harmonic function is
  \beq \chi(\rv)=\sum_i\log|\rv-\rv_i|.\eeq  
  Indeed, $\chi+{\rm i}\psi=\sum_i\log(\zeta-\zeta_i)$ is an analytic function. From the Cauchy-Riemann equations 
  \beq \nabla_r\chi=\nabla_\theta\varphi\quad, \quad \nabla_\theta \chi=-\nabla_r\varphi\eeq  
  it follows that
  \beq |(\nabla\varphi-\half\Omega\,r\,{\mathbf e}_\theta)|^2=|(\nabla\chi-\half\Omega\,r\,{\mathbf e}_r)|^2.\eeq  
The field
  \beq \mathbf E(\rv)\equiv \nabla\chi-\half\Omega\,r{\mathbf e}_r\eeq  
  can be interpreted as an electric field generated by unit \lq charges' at the positions of the vortices, $\rv_i$, and a uniform charge distribution of density {$\Omega/2\pi$}.
 
 We now distribute the vortices also with the same average density as the uniform charge. Thus every vortex $\rv_i$ sits at the center of  lattice cell $Q_i$ of area {$|Q_i|= 2\pi/\Omega$}, surrounded by a uniform charge distribution of the opposite sign so that the total charge in the cell is zero. If the cells were disc-shaped  Newton's theorem would imply that the \lq electric' field generated by the cell would vanish outside the cell, i.e, there would be no interaction between the cells. 
  Now the cells can not be disc shaped, but the closest approximation are \emph{hexagonal cells}, corresponding to a \emph{triangular lattice of vortices}, giving the optimal energy. The interaction between the cells, although not zero, is small because the cells have no dipole moment.

The energy of each cell, taking into account the factor $f(\rv)$ that cuts off the Coulomb field from the point charge (vortex) at a radius $t\ll\Omega^{-1/2}$, is 
\beq 2\pi \int_{t}^{\Omega^{-1/2}}(1/r)^2\, r\, dr +O(1)=\pi |\log (t^2\Omega)|+O(1).\eeq  
We now multiply by the number of cells, $\Omega/2$, and divide by the area, $\pi$, of the trap, obtaining
$\half\Omega |\log (t^2\Omega)| (1+o(1))$. The choice of the vortex radius $t$ is dictated by the error terms. The optimal choice is \beq t=\eps\cdot \rho^{-1/2}=\left\{
        	\begin{array}{ll}
            		\eps   &   \mbox{\rm if} \:\:\:\:  \Omega\lesssim 1/\eps \\
            		\mbox{} &   \mbox{} \\ (\eps/\Omega)^{1/2}
            		 &   \mbox{\rm if} \:\:\:\:  1/\eps \lesssim  \Omega\\ 
            		\mbox{} &   \mbox{} 
        	\end{array}
        	\right.
\eeq  
Indeed, for $\Omega\lesssim 1/\eps$ the average density $\rho$ is $O(1)$, while for 
$\Omega\gg 1/\eps$ the centrifugal forces cause a concentration of the GP wave function in a thin layer of thickness $O((\eps\Omega)^{-1})$ at the boundary and the average density is $O(\eps\Omega)$.
 
 A  lower bound matching the upper bound to subleading order can for
\beq\label{range} |\log\eps|\ll \Omega\ll 1/(\eps^2|\log\eps|)\eeq be proved making use of techniques from Ginzburg-Landau theory developed by S. Serfaty and E. Sandier \cite{SS1, SS2}.
 An important ingredient is a splitting of the domain into small cells and a rescaling  of lengths in each cell. This  allows to reduce the problem to a study of a Ginzburg-Landau functional in small cells with a new interaction parameter {$\epsilon$} and a magnetic field $h_{\rm ex}\ll |\log \epsilon|$. 
 
Outside the parameter range (\ref{range}) the ansatz (\ref{ansatz1}) does not give the correct energy to subleading order.
 In fact, for 
$\Omega\lesssim |\log \eps|$ there are only finitely many vortices 
(see, e.g., \cite{AftaDu}--\cite{RD}) and the upper bound
(\ref{bound2}) is too large. For $ 1/(\eps^2|\log\eps|)\lesssim \Omega$, on the other hand, a trial function different from (\ref{ansatz1}) gives lower energy than (\ref{bound1}). This is the trial function considered in \cite{CDY1} Eq.\ (3.36) for $\Omega\gg \eps^{-1}$ that corresponds to a {\it  giant vortex} \cite{FiB} where all the vorticity is concentrated at the center while the support of $\rho^{\rm TF}$ is vortex free. More explicitly, the trial function has the form
\beq\label{trial1} \psi(\rv)=\sqrt{\rho(r)}\exp({\rm i}[\Omega/2]\theta)\eeq
where $\rho(r)$ is a regularization of the  density  $\rho^{\rm TF}$ and $[\Omega/2]$ denotes the integer part of $\Omega/2$.
With this trial function the next correction to the TF energy is $O(1/\eps^2)$ that  is smaller than $\Omega |\log\eps|$ when $1/(\eps^2|\log\eps|)\lesssim \Omega$. This transition can be understood by the following heuristic argument, employing the electrostatic analogy: For $\Omega\gg 1/\eps$ the number of cells in the support of $\rho^{\rm TF}$ is of the order $1/\eps$.  Without vortices each cell has unit \lq charge', originating from the vector potential,  and the mutual interaction energy of the cells is of the order $1/\eps^2$. Putting a vortex in each cell neutralizes the charge so that the interaction energy becomes negligible, but instead there is an energy cost of order $\Omega|\log\eps|$ due to the vortices. Equating these two energies leads to $\Omega\sim 1/(\eps^2|\log\eps|)$ as the limiting rotational velocity above which the ansatz (\ref{ansatz1}) is definitely not optimal.

 It should be noted that also for $1/\eps\lesssim \Omega\ll1/(\eps^2|\log\eps|)$ one could for the upper bound replace the distribution of the vorticity on a lattice within the \lq hole'  by a single phase factor corresponding to a giant vortex at the origin, but in order to obtain the correction beyond the TF term the support of $\rho^{\rm TF}$ can not be vortex free. The detailed vortex distribution of the true minimizer of the GP energy functional is still an open question.

 \section{Techniques for Deriving the GP Equation}
This section contains a brief sketch of methods that have been used for a rigorous derivation of the three-dimensional GP energy functional from the many-body Hamiltonian.
 \subsection{Nonrotating trap}
 The first step is  a derivation of the ground state energy formula (\ref{gse}) in the case of a Bose gases in a box. For an upper bound \cite{dyson} (see also \cite{LSY1999, SeirNorw})  one can use a trial function of the form
 \beq\label{dysontrial} F(\xv_1,\dots,\xv_N)=F_2(\xv_1,\xv_2)\cdots F_N(\xv_1,\dots,\xv_N)\eeq  
 where 
\beq F_i(\xv_1,\dots,\xv_i)=f(\min_{j<i}|\xv_i-\xv_j|)\eeq  
 with a suitable function $f$ constructed from  the zero energy scattering solution. Although the trial function (\ref{dysontrial}) is not symmetric with respect to permutations of the variables it nevertheless gives an upper bound to the energy of the Bose gas, because the lowest energy of the Hamiltonian is obtained for a symmetric wave function, cf., e.g., \cite{liebs}.

  The lower bound \cite{LY1998} has the following main ingredients:\newline
   1) A lemma \cite{dyson} that allows to  replace a possibly \lq hard' interaction potential by a \lq soft', nearest neighbor  potential at the cost of increasing the range and sacrificing some kinetic energy.
\newline 2) Division of the  box into smaller boxes with Neumann boundary conditions. The size of the small boxes stays fixed in the thermodynamic limit.
\newline
   3) First order perturbation theory applied to the \lq soft', nearest neighbor  potential and an estimate of the error by Temple's inequality. 
   
The original proof of the lower bound \cite{LY1998} that was limited to nonnegative interaction potentials has recently been extended slightly to allow also potentials with a weak negative part \cite{lee}.
 
Consider next the case of an external trapping potential $V$ \cite{LSY1999, SeirNorw}.
  For an upper bound an adequate trial function has the form 
 \beq\label{trialgen} \Psi(\xv_1,\dots,\xv_N)=F(\xv_1,\dots,\xv_N) \prod_{j=1}^N\psi^{\rm GP}(\xv_j)\eeq  
 with $F$ as in (\ref{dysontrial}) and $\psi^{\rm GP}$ a minimizer of the  GP energy functional.
 
 For the lower bound one first uses the GP equation to replace $V$ by a potential depending on $\psi^{\rm GP}$. One approximates this potential by step functions in Neumann boxes and uses the lower bound derived  in \cite{LY1998} in each box.
 
 In the nonrotating case this technique can be easily extended to the TF limit, $g\to\infty$ but still $\rho a^3\to 0$ \cite{LSSY}.
 
\subsection{Rotating trap with $g$ and $\Omega$ fixed}\label{rotatingtrap}
  For the upper bound an apparent problem is that trial functions of the form (\ref{trialgen}) with nonsymmetric $F$ might lead to a lower energy than symmetric ones because the Hamiltonian is not real if $\Omega\neq 0$.
As noted in \cite{Seir2}, however, functions of the form (\ref{trialgen}) give an upper bound to the bosonic ground state energy  as long as 
 $F$ is a \emph{real} function.  In fact, using the GP equation for $\psi^{\rm GP}$  one has in this case 
 \begin{eqnarray}\label{real}\langle\Psi,H\Psi\rangle&=&NE^{\rm GP}+4\pi gN\int|\psi^{\rm GP}_i|^4\nonumber \\&+&\int\left(\sum_i|\nabla_i F|^2-8\pi g|\psi^{\rm GP}_i|^2F|^2+\sum_{i<j}v_{ij}|F|^2\right)\prod_i\psi^{\rm GP}_i\end{eqnarray}
 with $\psi^{\rm GP}_i\equiv\psi^{\rm GP}(\xv_i)$ and $v_{ij}\equiv v(\xv_i-\xv_j)$. Since the right hand side of (\ref{real}) is a real quadratic form it does not matter whether the infimum is taken over symmetric or nonsymmetric  $F$ and a trial function of the form (\ref{dysontrial}) can be used.

  For a lower bound the lemma of \cite{dyson} is still important to \lq soften' the interaction potential, but the division into Neumann boxes fails and an entirely different technique technique is required. This long standing problem was solved in \cite{Lieb1} by using coherent states.  See also \cite{Seir3} for a r\'esum\'e. 
 
 There remain two cases, however, that are not covered by the  proof in \cite{Lieb1} that is restricted to fixed $g$ and $\Omega$:
 \begin{itemize}
 \item The TF limit, $g\to\infty$ and/or $\Omega\to\infty$,  in rotating, anharmonic traps.
 \item The limit $\Omega\to\Omega_c$ in harmonic traps, where a GP description may completely break down.
 \end{itemize}

 \section{The TF Limit of the Many-Body Ground State Energy in Anharmonic Traps}

This section is based on  \cite{BCPY} and concerns the leading order contribution to the many-body ground state energy of a dilute,  three-dimensional Bose gas in the TF limit $g\to\infty$.  The rotational velocity may at the same time also tend to infinity.  The corresponding problem for the two-dimensional GP energy was discussed in Section 4 where also the subleading correction to the leading TF contribution could be evaluated exactly. In the many-body context the extraction of the leading order contribution is already a non-trivial problem. A proof that the GP energy  gives also the subleading contribution to the many-body energy for rotating gases if $g\to\infty$ has still to  be completed.
 
We recall from (\ref{gpfunc}) the GP energy functional 
\begin{equation}
    \mathcal{E}^{\mathrm{GP}}_{g,\Omega}\left[ \psi \right] = \int_{\R^3}\left\{|({\rm i}\nabla+\,{\mathbf A})\psi|^2+(V-\hbox{$\frac 14$}\Omega^2r^2)|\psi|^2+
g|\psi|^4\right\}\hbox{\rm d}^3{\mathbf x},   \label{GP functional}
\end{equation}
where the dependence on the parameters $g$ and $\Omega$ has been explicitly indicated. The confining, external potential $V$ is assumed to be smooth and satisfy 
 $V(\xv)>0$ for $|\xv|=1$ and $V(\lambda \xv)=\lambda ^sV(\xv)$ for all $\lambda>0$ with some $s>2$.

The GP energy is
\begin{equation}
E_{g,\Omega }^{\mathrm{GP}}\equiv {\inf }\left\{\mathcal{E}_{g,\Omega}^{\mathrm{GP}}\left[ \psi
\right]:\; \Vert \psi \Vert _{2}=1 \right\}.   \label{GP energy}
\end{equation}
The infimum is, in fact, a minimum and we denote  any normalized  minimizer by $\psi^{\rm GP}_{g, \Omega}$.  The corresponding density is 
$\rho^{\rm GP}_{g, \Omega}=|\psi^{\rm GP}_{g, \Omega}|^2$.

The TF energy functional, obtained from the GP functional by dropping the kinetic term, is 
\begin{equation}\label{tffunctional}
    \tff\left[ \rho \right] \equiv \int_{\RT}\left\{ V\rho -\frac{1}{4}\Omega ^{2}r^{2}\rho +g\rho
^{2}\right\}\diff^3
\xv. 
\end{equation}
It is defined for nonnegative densities $\rho(\cdot)$, and the TF energy 
\beq
    E_{g,\Omega }^{\mathrm{TF}}\equiv {\inf }\left\{\tff \left[ \rho
\right]:\;  \Vert\rho\Vert _{1}=1 \right\}.  \label{TF energy}\eeq
is attained for the
unique minimizer:
\begin{equation}
\rho _{g,\Omega }^{\mathrm{TF}}\left( {\mathbf x}\right)
=\frac{1}{2g}\left[
\mu _{g,\Omega }^{\mathrm{TF}}+\frac{1}{4}\Omega ^{2}r^{2}-V\left( {\mathbf x}
\right) \right] _{+}, \label{TF_ density}
\end{equation}
where $[\cdot]_+$ denotes the positive part and $\mu _{g,\Omega }^{\mathrm{TF}}$ is the TF chemical
potential determined by the normalization $||\rho _{g,\Omega }^{\mathrm{TF}}||_{1}=1$. 

The TF energy and density satisfy some simple \emph{scaling relations} that are useful for distinguishing different parameter regimes. 
Writing
$\xv=\lambda\xvp$ and $\rho(\xv)=\lambda^{-3}\rho'(\xvp)$ one obtains
\begin{eqnarray}
       \mathcal{E}^{\mathrm{TF}}_{g,\Omega} \left[ \rho \right] = \lambda^{-2} \int_{\RT}  \left\{ \lambda^{s+2} V(\xvp)\rho' -\hbox{$ \frac{1}{4}$}(\lambda^2 \Omega)^{2}{r^{\prime}}^{2} \rho' + g \lambda ^{-1}(\rho')^2\right\}\diff^3\xv.
\end{eqnarray}
We now distinguish two cases:\\

\noindent{\bf  A.} Rotational effects are at most comparable to the interaction effects:  We equate $\lambda^{s+2}$ with $g\lambda^{-1}$, i.e., choose
\beq\lambda=g^{1/(s+3)}\label{lambda1}\eeq
and obtain
\begin{equation}
g^{-{s}/(s+3)%
}E_{g,\Omega }^{\mathrm{TF}} =E_{1,\omega }^{\mathrm{TF}}
\label{TF big interaction}
\end{equation}
with \beq \omega\equiv  g^{-{(s-2)}/{2\left( s+3\right) }}\Omega\label{omega}
\eeq 
and likewise \beq g^{{3}/(s+3)}\rho _{g,\Omega
}^{\mathrm{TF}}\left( g^{{1}/(s+3)}{\mathbf x}\right)=\rho
_{1,\omega }^{\mathrm{TF}}\left(
{\mathbf x}\right).\label{TFdensityscaling} \eeq 
Hence  the TF theory has one parameter, $\omega$. The case $\omega=0$
corresponds the TF functional without rotation.\\

\noindent {\bf B.} If $\omega\to \infty$ the rotational term completely dominates the interaction term. In this case it is appropriate to use a different scaling and equate 
$\lambda^{s+2}$  with $(\lambda^2\Omega)^2$, i.e, take
\beq\lambda=\Omega^{2/(s-2)}.\eeq
We then obtain 
\begin{equation}
\Omega^{-2s/(s-2)}E_{g,\Omega }^{\mathrm{TF}} =E_{\gamma,1}^{\mathrm{TF}}
\label{TFenergyscalingstrong}
\end{equation}
with \beq\label{gamma} \gamma\equiv  \Omega^{-2(s+3)/(s-2)}g=\omega^{-2(s+3)/(s-2)}.
\eeq

Moreover, as $\omega\to \infty$, i.e., $\gamma\to 0$, we have
\begin{equation}
\lim_{\gamma\to 0 }E_{\gamma,1}^{\mathrm{TF}} =E_{0,1}^{\mathrm{TF}}=%
{\inf }\left\{
V(\xv)-\hbox{$\frac{1}{4}$}r^{2}:\; {\xv \in \mathbb{R}^{3}}\right\}<0  \label{TF strong rotation}
\end{equation}
while $\rho _{\gamma,1
}^{\mathrm{TF}}$ converges
 to a measure supported on the set {}{$\mathcal{M}$ }of
minima of the function $W(\vec x)\equiv V(\xv)-\hbox{$\frac{1}{4}$}r^{2}$.  As a useful fact we note that  all points in $\mathcal M$ have the same distance from the rotation axis.

\noindent Examples:\\
\smallskip
If $V(\vec x)=ar^s+b|z|^s$, then $\mathcal M$ is  a circle in the $z=0$ plane. \\
\smallskip
If  $V(\vec x)=a|x|^s+b|y|^s+c|z|^s$ with $a\neq b$ then $\mathcal M$ consists of two points.

The scaling properties of the TF theory already suggest that one should distinguish between the following three cases:\begin{itemize}
\item {\bf Slow rotation}, {$\omega\ll 1$}: The effect of the rotation is negligible to leading order.
\item {\bf Rapid rotation}, {$\omega\sim 1$}: Rotational effects are comparable to those of the interactions.
\item {\bf Ultrarapid rotation}, {$\omega\gg 1$}: Rotational effects dominate.
\end{itemize}
The main result is that in all cases the TF energy, suitably scaled, is the leading term of the many-body ground state energy $E_{g,\Omega }^{\mathrm{QM}}$  as $N\to\infty$ and $g\to\infty$, provided the gas stays \emph{dilute}, i.e., condition (\ref{dilute}) is fulfilled. As a measure for the density one can use $N\, \Vert \rho_{g,\Omega }^{\mathrm{TF}}\Vert_{\infty}$, so with $a\sim g/N$ the diluteness condition can be written as  $N^{-2}g^3\Vert \rho_{g,\Omega }^{\mathrm{TF}}\Vert_{\infty}\ll 1$.

\begin{theorem}[QM Energy Asymptotics]
\label{TF theorem 1} 

 Assume that
$N^{-2}g^3\Vert \rho_{g,\Omega }^{\mathrm{TF}}\Vert_{\infty}\to 0$ as $N\to\infty$.

\noindent\hskip .5cm{\rm (i)} If $g \to\infty$  and  $\omega\to 0$ as $N\to\infty$, then {}{\beq  {\lim_{N\to\infty} }\left\{ g^{-{s}/(s+3)}N^{-1} E_{g,\Omega }^{\mathrm{QM}}\left( N\right)  \right\}
=E^{\rm TF}_{1,0}\eeq  }
\noindent\hskip .5cm{\rm (ii)} If $g\to\infty$  and $\omega>0$ is fixed as $N\to\infty$, then \beq {\lim_{N\to\infty} }\left\{ g^{-{s}/(s+3)}N^{-1} E_{g,\Omega }^{\mathrm{QM}}\left( N\right)  \right\}
=\ttfe\eeq  
\noindent\hskip .5cm{\rm (iii)} If $\Omega\to \infty$ and
$\omega\to\infty $ as $N\to\infty$, then
 \beq \lim_{N\to\infty} \left\{\Omega^{- {2s}/(s-2)}N^{-1} E_{g,\Omega
}^{\mathrm{QM}}\left( N\right)
 \right\} =\tfeinf.\eeq
 \end{theorem}

The energy estimates also lead to limit theorems for the particle density $\rho _{N,g,\Omega
}^{\mathrm{QM}}$  in a many-body ground state:

\begin{theorem}[QM Density Asymptotics for $\omega<\infty$]
    \label{TF theorem 2}
    \mbox{} \\
   Under the conditions of Theorem 4 (i) and (ii) we have \beq g^{{3}/(s+3)}N^{-1}\rho _{N,g,\Omega
}^{\mathrm{QM}}\left( g^{{1}/(s+3)}{\xv}\right)\to\rho
_{1,\omega }^{\mathrm{TF}}\left(
{\xv}\right)\label{QMdensityscaling} \eeq in weak $L_1$ sense.
\end{theorem}

\begin{theorem}[QM Density Asymptotics for $\omega\to \infty$]
    \label{TF theorem 3}
    \mbox{} \\
    Under the conditions of Theorem 4
(iii), the scaled particle density \beq\Omega^{ 6/(s-2)}N^{-1}\rho
_{N,g,\Omega }^{\mathrm{QM}}\left( \Omega^{
2/(s-2)}{\xv}\right)\eeq converges to a  probability measure with support in
$\mathcal{M}$.
\end{theorem}

For the proof of Theorem \ref{TF theorem 1}  it is necessary to prove first a corresponding result for the GP theory:
\begin{theorem}[GP Energy Asymptotics for $\omega<\infty$]
       \mbox{} \\
For  $g\to\infty$ with $\omega$ fixed we have
    \beq
        g^{-{s}/(s+3)} \gpe = \ttfe + O \left( g^{-(s+2)/{2(s+3)}} \log g \right).
    \eeq
\end{theorem}
This theorem is  a partial extension of Theorem \ref{thm1} to 3D and general homogeneous potentials but since it only estimates the order of the subleading term in the energy and does not give not its precise form, its proof is simpler. In fact, the lower bound is obtained simply by 
dropping the nonnegative kinetic energy term
$|({\rm i}\nabla+{\mathbf A})\psi|^2$ from $\gpf$. The upper bound  is proved by using a  (suitably  scaled) three-dimensional version of the trial function (\ref{trial1}). In particular, this trial function exhibits a lattice of vortex lines that partly compensate the contribution of the vector potential to the kinetic energy term.

In the case of ultrarapid rotations, $\omega\to\infty$, we have\\
\begin{theorem}[GP Energy Asymptotics for $\omega\to \infty$]
\label{EnAsOmegainfty} \mbox{} \newline {\it If $\Omega$ and $\omega\to\infty$ then 
\beq
\Omega ^{-{2s}/(s-2)}\gpe =
\tfeinf+{O}\left( \Omega ^{\prime -1}+\gamma^{
{2}/{5}}\right) , \eeq} where  $\Omega ^{\prime }=
\Omega^{(s+2)/(s-2)}$ and $\gamma=\omega
^{-2(s+3)/(s-2 )}.$
\end{theorem}
This can be proved by picking a point $\xv_0\in{\mathcal M}$ where $V-r^2/4$ is minimal and using a trial function of the form
\beq
\label{trial function 2} \tilde\psi(\xv) = \sqrt{h_{\delta }({\xv}-{\xv}_{0})} \exp \left\{ i\Omega {\xv}\cdot \left( {\mathbf e}_{z}\wedge {\xv}%
_{0}\right) /2\right\}  \eeq
where $h_\delta$ is a smooth approximation of the delta function.

We now discuss briefly the main steps in the proof of Theorem \ref{TF theorem 1}.\smallskip

\noindent \emph{1. Upper bound.} \\ 
For slow to rapid rotation ($\omega<\infty$) one can use a trial function of the form
(\ref{trialgen}) 
with a {\it real} function $F$ as discussed in Subsection \ref{rotatingtrap}.
The bound on $E^{\rm GP}$ in terms of $E^{\rm TF}$ was already described; it leads also to a bound on the GP density in terms of the TF density.

For ultrarapid rotations, $\omega\to\infty$,  the trial function has to be modified. One takes
\beq \Psi(\xv_1,\dots,\xv_N)=F(\xv_1,\dots,\xv_N)\prod_{i=1}^N\psi(\xv_i)\eeq  
with $F$ as in (\ref{dysontrial}) and \beq \psi(\xv)=\sqrt{\rho(\xv)}\exp(iS(\xv))\eeq  
where $\rho$ is a regularized TF density and the phase factor
$S$ corresponds to a \lq giant vortex'  centered at the origin:
\beq S(\xv)=\left [\hbox{$\frac12$}r^2_\Omega \Omega\right]\theta\eeq  
where $[\cdot]$ denotes the integer part and 
$r_\Omega$ is the radius of the set $\mathcal M_\Omega$ where $V-\Omega^2r^2/r$ is minimal. (This set is always a subset of a cylinder.)\\
 
\noindent \emph{2. Lower bound.}

\noindent In the case $\omega<\infty$ one first uses the \emph{diamagnetic inequality}
\beq |(\nabla-{\rm i}\,{\mathbf A})\psi|^2\geq |\nabla \psi|^2.\eeq  
Then  one writes 
\beq V-\frac{\omega^2}4 r^2= V-\frac{\omega^2}4 r^2-\mu+\mu\geq -2\rho^{\rm TF}+\mu\eeq  
with the TF chemical potential $\mu=\mu^{\rm TF}_{1,\omega}$ that satisfies
\beq \mu=E^{\rm TF}+\int(\rho^{\rm TF})^2. \eeq  
 We then have  to bound the Hamiltonian
\beq \tilde H=\sum_i(-\Delta_i-2g'\rho^{\rm TF}_i)+\sum_{i<j}v_{ij}'\eeq
where $g'$ and $v'$ are suitably scaled versions of the coupling constant and the interaction potential respectively.  This is done by introducing Neumann boxes where $\rho^{\rm TF}$ is approximately constant and using
the basic bound of \cite{LY1998} for the ground state energy in each box,
\beq E_0(n,\ell)\geq 4\pi a(n/\ell^3)(1-o(1)),\eeq  
where $n$ is the particle number, $\ell$ the box size and $o(1)\to 0$ if $a n/\ell^3\to 0$ and $n\to\infty$.
This leads to the lower bound 
\beq -g'\int (\rho^{\rm TF})^2(1+o(1))\eeq   for $\tilde H$ and altogether the bound \beq NE^{\rm TF}(1-o(1))\eeq   for the energy.

{}

The case of ultrarapid rotations, $\omega\to\infty$,  is simpler than the case of finite $\omega$ since for the lower bound for the energy we can ignore the (positive) interaction altogether. Namely, dropping  the positive kinetic term,
we can  write for any $\Psi$
\begin{equation}
 \Omega ^{-{2s}/(s-2)}N^{-1}\left\langle
\Psi,H\Psi\right\rangle \geq C_{\Psi}+{\inf_{\RT}}\,
W
\end{equation}
with $W(\xv) =V( \xv) -r^{2}/4$
and
\begin{equation}
C_{\Psi} = \dint_{\RT}{\rho'}^{\rm TF}(\xv{%
})\left( W({\xv})-{\inf_{\RT} }\,W\right)
\mathrm{d}^3{\xv}
+ \Omega ^{-{2s}/(s-2)}N^{-1}{\sum_{1\leq i<j\leq N}
}\left\langle \Psi,v_{ij}\Psi\right\rangle .
\end{equation}
Since the interaction potential $v$ is by assumption nonnegative the same holds for
$C_{\Psi}$, and $ \inf W=\tfeinf$.

\section{Remarks on Rapid Rotation in Harmonic Traps}\label{harmonic}

In a  quadratic trapping potentials, $V(\xv)=\half \Omega_{\rm osc}^2 |\xv|^2$, interesting effects are expected to occur when $\Omega$ approaches  the critical frequency
\beq \Omega_c=\sqrt 2 \,  \Omega_{\rm osc}\eeq
 at which \beq V_{\rm eff}(\xv)\equiv V(\xv)-\hbox{$\frac 14$}\Omega^2r^2\eeq is no longer bounded below \cite{Ho}, see also \cite{FetterRMP} for a review. As  $\Omega_c-\Omega\to 0_+$ the effective radius $R$ of the condensate tends to infinity  and the system becomes effectively two-dimensional. Equating the potential energy corresponding to $V_{\rm eff}$ and the interaction energy leads to a simple estimate for the radius: 
\beq R\sim \left(\frac {Na}{(\Omega_c-\Omega)\Omega_c^{1/2}}\right)^{1/4}.\eeq
The expected number of vortices is $\sim \Omega R^2$ so the ratio between the particle number and the number of vortices is
\beq\frac{\#\,\hbox{\rm particles}}{\#\,\hbox{\rm vortices}}\sim \frac N {\Omega R^2}\sim \big(N(1-\Omega/\Omega_c)/a\Omega_c^{1/2}\big)^{1/2}.\eeq
The expectations are now:
 \begin{itemize}
\item If  $N(1-\Omega/\Omega_c)/a\Omega_c^{1/2}\to\infty$, the many-body ground state is still well described by GP theory.
 \item  If $N(1-\Omega/\Omega_c)/a\Omega_c^{1/2}$ stays small, the many-body ground state is highly correlated and there is no GP description. The wave function should be well approximated by a function of Laughlin type in the variables $\zeta=x+{\rm i}y$:
\beq \Psi(\zeta_1,\dots,\zeta_N) \sim\prod_{i<j}(\zeta_i-\zeta_j)^2\prod_i\exp(-\Omega |\zeta_i|^2/4)\eeq
 corresponding to a \emph{Fractional Quantum Hall Effect (FQHE)} in the lowest Landau level.
 \end{itemize}
  So far, neither of these expectations has been rigorously proved.

\section{Conclusions}
For rapidly rotating, dilute Bose gases in anharmonic traps the leading energy and density asymptotics of the many-body ground state as the coupling parameter $g\sim Na$ tends to infinity can be calculated  from the simple density functional (\ref{tffunctional}) for arbitrary rotational velocities. Moreover, within two-dimensional Gross-Pitaeveskii theory also the subleading terms in the energy, corresponding to a lattice of vortices, can be evaluated exactly. The following problems are still open:
\begin{itemize}
\item Prove that GP theory provides the subleading term of the many-body ground state energy for rapidly rotating, anharmonic  traps in the TF limit.
\item For harmonic traps, prove that GP theory still applies in the limit $\Omega\to\Omega_c$, provided $N(1-\Omega/\Omega_c)/a\Omega_c^{1/2}\to\infty$.
\item Understand rigorously the transition into the FQHE regime if $N(1-\Omega/\Omega_c)/a\Omega_c^{1/2}$ remains finite. \end{itemize}

\section*{Acknowledgments} The work reported here was supported by the Austrian Science Fund (FWF) grant P17176-N02, the EU Post Doctoral Training Network HPRN-CT-2002-00277 ``Analysis and Quantum'' and the ESF Research Networking Programme INSTANS. 
Hospitality at the  Institute Henri Poincar\'e, Paris, the Niels Bohr International Academy, Copenhagen, and the Science Institute, University of Iceland, Reykjavik  is also gratefully acknowledged.

\end{document}